\documentstyle[epsfig,prl,floats,aps]{revtex}
         \newcommand{\be}{\begin{equation}}
         \newcommand{\ee}{\end{equation}}
         
\begin{document}
         \renewcommand{\topfraction}{0.99}
         \renewcommand{\bottomfraction}{0.99}
         \twocolumn[\hsize\textwidth\columnwidth\hsize\csname
         @twocolumnfalse\endcsname

         \title{Inflation From $D-\bar{D}$ Brane
         Annihilation}
         \author{Stephon H.S. Alexander}
         \footnote{E-mail: stephon@ic.ac.uk}
         \address{~\\
         Theoretical Physics Division, Imperial
         College, The Blackett Laboratory Prince Consort Rd., London, U.K. SW72BZ}

         \maketitle

         \begin{abstract}
         We demonstrate that the initial conditions for inflation are met when
         a $D5-\bar{D}5$ brane annihilate. This scenario uses Sen's
         conjecture that a co-dimension two vortex forms on the worldvolume of
         the annihilated 5-brane system. Analogous to a ``Big Bang'', when the
         five branes annihilate, a vortex localized on a 3-brane forms and its
         false vacuum energy generates an inflationy space-time.
         We also provide a natural mechanism for ending inflation via the
         motion of the vortex in the bulk due to its extrinsic curvature. We
         also suggest a consistent way to end inflation and localize matter on
         our space-time.
         \end{abstract}

         ]

         \section{Introduction}

         It has been suggested by Rubakov and Shaposhnikov
         and later by other investigators that our universe may be a defect
         embedded in a higher
         dimensional bulk spacetime\cite{rubakov}. This idea has reemerged in a
         more concrete
         context, namely to solve the gauge hierarchy problem and localize 4-D
         gravity as well as the matter fields of the Standard
         Model\cite{lisa,dvali}. It was
         demonstrated that the RS scenario still exhibits a flatness problem
         and neccesitates an inflationary epoch\cite{rocky}. Moreover, recent
         observations of the CMB agree with the inflationary scenario which
         resloves the problems of the standard big bang model (SBB) such as the
         Horizon
         , Flatness and Formation of Structure problems.

         Nonetheless, so far most brane-world
         descriptions are constructed from the bottom-up necessitating other forms
         of fine tuning.
         In light of the flatness, structure formation problems and especially
         the trans-plankian problems, we expect
         quantum gravitational effects to become important in the early
         universe \cite{rhb1}.

         In light of the limitations of the effective field theories applied
         to inflationary scenarios, inflation should arise
         as a prediction from string theory, since string theory
         incorporates natural ways of resolving curvature singularities and
         field theory divergences via.
         and S and T-dualities\cite{easson}.
         Nonetheless, inflation requires very special initial conditions that
         are usually relinquished to the specifics of an effective theory
         \cite{rhb2}. In this paper,
         we investigate the non-BPS sector of superstring theory and show that
         the initial conditions for inflation are realized quite naturally.
         We will show that when two five branes annihilate, an inflating three
         dimensional hypersurface will emerge as a result.

         A key to realizing inflation from D-branes is the fact that
         they are space-time topological defects. It has been appreciated for a
         while that topological defects play an
         important role in the early universe. Indeed, Vilenkin and Linde
         demonstrates that if the symmetry breaking scale associated with the
         formation of a defect is on the order of the planck scale, a topological
         defect will drive inflation free of the fine tuning problems which
         usually plague inflationary scenarios; hence, inflation becomes an
         issue of topology \cite{vilen,linde}.

         Therefore, the idea for D-brane driven inflation is quite simple. When the
         branes annihilate a co
         dimension 2 vortex forms at the center of the created three brane,
         identified as our space-time[see figure 1]. For an observer in this spacetime, the
         vacuum energy dominates and generates eternal inflation. When
         inflation begins, it makes the tachyon (inflaton) homogenous and very
         close to zero near the core of the three brane vortex. Outside the
         core of the three brane, inflation will eventually end and such an
         observer
         will see a black 3-brane. Compared to
         the hubble length scale these two observers are exponentially
         separated, so there will be no cosmological problems.
	 \begin{figure}
         \centerline{\epsfxsize=3in\epsffile{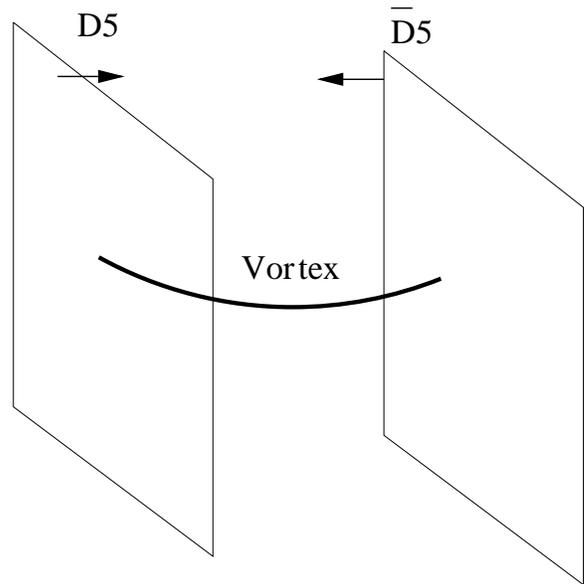}}
         \caption{This is a depiction of the annihilation of the
	$D5-\bar{D}5$ brane system, represented as a two dimensional
	surface.  A vortex, represented as a line, forms after the D5 branes
	annhihlate. In our mechanism, this vortex initiates an inflating
	3+1D hypersurface.}
         \label{D5}

         \end{figure}

         In order to discuss D-brane inflation it is necessary to
         provide an
         analysis of non-Bps D-branes. In section[II] and [III] we review generic
         properties
         of Bps and non-Bps Dbranes to set the stage for its cosmological
         relevance. In section [IV] we will then review in general the
         realization of defect driven inflation. In Section [V] we extend the
         analysis of
         defect-driven inflation to non-bps D-brane systems. An explicit analysis
         and solution of the effective tachyon vortex field theory coupled to
         gravity is
         covered in section [VI]. A discussion of
         how inflation can end and matter is localized is discussed in section
         [VII]. We conclude
         with
         some open issues in light of D-brane physics and string theory.

         \section{Generic Properties of D-Branes}

         At long wavelengths, $\lambda > l_{s}$ the dynamics of a Dp-brane is
         well described by the sum of the Dirac-Born-Infeld (DBI) action and a
         Wess-Zumino (WZ) term,
         \be S_{p} = S^{DBI}_{p} + S^{WZ}_{p} \ee
         The DBI term being
         \begin{eqnarray} && S_{p}^{DBI}\nonumber \\
	 && =T_{(p)}\int d^{p+1}x \sqrt{det[(G_{\mu\nu} +
         B_{\mu\nu})\partial_{m}Y^{\mu}\partial_{n}Y^{\nu} +
         2\pi \alpha' F_{mn}]} \end{eqnarray}
         Where $F_{mn}$ is the world-volume Born-Infeld field 
         strength, $B_{\mu\nu}$ is the bulk antisymmetric field, 
         $Y^{\mu\nu}$ are the collective coordinates which describe
	oscillations transverse to the world volume the D-p Brane,
         $G_{\mu\nu}$ is the target-space metric.  The tension of the 
         Dp-brane $T_{(p)}$ is,
         \be T_{(p)} =2\pi(4\pi^{2}\alpha^{'})^{-(1+p)/2} e^{-\phi} \ee
	 where $\alpha' = l_{s}^{2}$.
         Our conventions are the ten-dimensional target spacetime vectors
         labeled
         by $\mu=0,...,9$. The world-volume directions are
         $m,n,...=0,...,p$, while the directions transverse to the Dp-brane
         will
         be labeled $a,b,..=p+1,...,9$. The low energy limit of the WZ term in
         the action [DBI] can be deduced by requiring the absence of chiral
         anomalies in an arbitrary configuration of intersecting D-branes.

         All superstring theories admit a myriad of Dp-brane species.
         Nonetheless, there are some features that are generic among all
         D-branes
         which we aim to exploit in cosmology.
         The most outstanding generic physical feature of D-branes
         are their low energy, long wavelength behavior.
         The transverse fluctuations of the D-brane is concretely described by
         the 9-p+1 scalar fields (more geometrically speaking, the
         Normal-Bundle). An observer on the brane will see these scalars as
         the D-term in the corresponding Super-Yang-Mills theory
         which reside on the the D-brane's world volume (WV). Similarly, gauge
         fields residing in
         the D-brane's WV describe the longitudinal fluctuation of the
         D-brane.

         Notice that the above action describes a supersymmetric D-brane. As
         a result, the dynamics of this brane is constrained to locally
         supersymmetric gravitational backgrounds. Therefore cosmological
         space-times including De-Sitter is inadmissible as they will break
         supersymmetry on the D3-brane worldvolume. Hence our brane is
         necessarily non-bps in order to incorporate dynamical gravity. We
         are led to therefore consider the evolution of a non-bps 3-Brane
         cosmology. Even in the early universe any brane world scenario will
         have to incorporate the non-bps sector of string theory, hence
         non-bps D-branes. But, how do non-bps branes arise from first
         principles? In particular, we are interested
         in the cosmological implication of D-brane anti-D brane annihilation
         since it has been conjectured by Sen that this state is equivalent to
         vacuum.

         \section{Non-Bps D-brane}
         In this section we provide a first principle approach to non-bps
         D-brane that will be compatible with early universe cosmology. One
         first needs to understand from a stringy perspective how non-Bps
         D-branes arise and evolve.

         Similar to point particles, when a D-brane and an Anti D-brane are
         coincident they will annihilate. However, unlike point particles,
         the D-brane annihilation process is more involved since each of these
         D-branes have a U(1) gauge field theory living on its world volume.
         Therefore, the fate of these gauge fields during
         and after the annihilation process play a crucial role in
         determining decay product.  

         When the branes are coincident a tachyonic instability sets in.
 	The open string connecting the $D-\bar{D}$ brane has a spectrum arising
from a GSO projection, $(-1)^{F}$, that is the reverse of the usual
one. \footnote{For a nice review, read \cite{john,sen3}}  Usually, the GSO 
         projection acts to get rid of the tachyon in the open strings 
         which end on a Dp-Brane.  However, for a $D-\bar{D}$ string 
         the tachyon still survives despite the GSO projection.  The
	tachyonic instability signals the eventual annihliation of the
	coincident brane and anti-brane.  Sen conjectured that the
         Tensions of the D-anti D
         brane pair and the negative potential energy of the tachyon is
         exactly zero\cite{sen1,sen2}.
         \be 2T_{D} + V(T_{0})=0 \ee
         where $T_{D}$ is the tension of the D-brane and $V(T_{0})$ is the
         value of the tachyonic potential at its minimum. Therefore as the
         branes annihilate the tachyonic fields evolve towards a true vacuum
         where the tensions of both branes are equivalent to the minimum of
         the tachyonic potential. In this case the branes will annihilate to
         the the closed string vacuum and there will be no remnant branes.
         We are of course, interested in the case when a lower dimensional
         brane is created as a by product of the annihilation process. The
         corresponding equation describing this process is:
         \be 2T_{D_{p+2}} + V(T_{0}) - T_{D_{p}}=0 \ee
         where $T_{D_{p}}$ is the tension of the lower dimensional brane which
         the higher dimensional branes annihilate into.  In this 
         case, there is excess energy density associated with the 
         $Dp$ brane which is created in the annihilation process.

         Let us look at a concrete case of a vortex confugration on a
         membrane-antimembrane pair in type IIA string theory. Here, the
         tachyon associated with the open string ending on the membrane and the
         anti-membrane is a complex scalar field $T$. There is a
         $U(1)\times U(1)$ gauge field living on the world volume of the
         membrane anti-membrane system. The tachyon carries one unit of
         winding charge under these gauve fields. Let $A_{\mu}^{1}$ and
         $A_{\mu}^{2}$ denote the gauge fields arising from the D2-brane and
         the anti-D2-brane respectively. The resulting kinetic term for the
         tachyon is:

         \be |D_{\mu}T|^{2} \ee,
         where
         \be D_{\mu}T=(\partial - iA_{\mu}^{1} + iA_{\mu}^{2})T. \ee

         The form of the perturbative tachyonic potential we employ is
         \cite{kraus}
         \be V(T) = (|T|^{2}- m^{2})^{2} \label{tpot} \ee

         The general static, finite energy vortex like configuration for the
         tachyon field described in the polar coordinates on the membrane
         world volume in the asymptotic regime takes the form

         \be T \simeq T_{0}e^{in\theta}\ee
         \be A_{\theta}^{1} - A_{\theta}^{2}
         \simeq 1 {\rm , as }  \rightarrow  \infty . \ee
        
	 Hence as $r \rightarrow \infty $, both the kinetic and potential
         energy will vanish rapidly.  This defect is
         identified as a stable, finite mass particle in type IIA string
         theory.
         This particle carries one unit of magnetic flux associated with the
         gauge field on the world volume of the membrane anti-membrane system.
         \be \oint (A^{1} - A^{2}).dl =2\pi \ee
         Hence, this nontrivial flux implies that the particle carries one
         unit of $D0$ brane charge\cite{sen3}. It has been argued using boundary
         conformal field theory calculations that this soliton is
         indistinguishable from a $D0$ brane. The above construction can be
         trivially generalized to represent the p-brane of type II string
         theory as a vortex solution on the $(p+2)$-brane anti-$(p+2)$-brane
         pair\cite{sen3}. We are of course interested in the case of p=3, the
         D3-brane.

         We are interested in knowing the cosmological implications and
         consequences of the
         identification of the vortex-defect as a co-dimension 2 D-brane after a
         brane and an anti-brane annihilates.

         \section{Inflationary Mechanism: The Defect Solution}

         It is astonishing that the core of topoligical defects can undergo
         cosmic inflation without the need of fine tuning. Both Linde and
         Vilenkin first calculated the criterion for a defect core to
         undergo inflation. To make our analysis clear let us consider
         inflation of a domain wall. The Lagrangian of a domain wall is:

         \be L= \frac{1}{2}(\partial_{\mu}\phi)^{2} -
         \frac{\lambda}{4}(\phi^{2} - \frac{m^{2}}{\lambda})^{2}\ee,
         where $\phi$ is a real scalar field. The Lagrangian possesses a
         $Z_{2}$ symmetry that is spontaneously broken and hence domains are
         formed with $\phi= \pm \eta$ where $\eta = \frac{m}{\sqrt{\lambda}}$.
         These domains are divided by kinks (domain walls) which interpolate
         between the two minima. The domain wall configuration is represented
         as

         \be \phi= \eta \tanh(\sqrt{\frac{\lambda}{2}} \eta x). \ee

         What are the conditions for a universe separated into two domains by
         a domain wall to inflate? To answer this question we need
         to show that there is a regime in parameter space of the domain wall
         coupling and symmtery breaking scale that will yield an exponentially
         expanding space-time background.

         We first need to find the thickness of the domain wall in
         flat space-time, which is obtained by balancing the gradient and
         potential energies at the core of the wall. The potential energy
         strives to keep the domain wall field configuration on the vacuum
         manifold, hence minimizes the DW thickness, while the gradient energy
         provides tension to spread out the wall thickness. The potential
         energy density of the wall is obtained by evaluating the potential
         at $\phi=0$ since the field configuration is localized in the core of
         the wall. This gives $\rho_{d} = \lambda \eta^{4}$.
	 
         The wall thickness, $\delta_{0}$ in flat spacetime is determined by the
         balance of
         the gradient and potential energy, $(\frac{\eta}{\delta_{0}})^{2} \sim
         V_{0}=V(\phi=0) $ Hence,

         \be \delta_{0} \simeq \eta (V_{0}^{1/2}) \ee
	 From the Friedmann equation the horizon size corresponding to the
         vacuum energy $V_{0}$ in the iinterior of the wall is

         \be H_{0}^{-1} = M_{p}(\frac{3}{8\pi V_{0}})^{1/2} \ee
         where $M_{p}$ is the Planck mass.

         If $\delta_{0} << H^{-1}$, then gravity will not affect affect the
         wall structure in the transverse direction, hence, we do not expect the
         wall thickness to change. However, for $\delta_{0} \geq H_{0}^{-1}$ the
         size of of the false vacuum region inside the wall is greater than
         $H_{0}^{-1}$ in all three directions, and according to the Einstein
         field equations this region will undergo inflationary expansion.
         Furthermore, using the above two conditions, we find that inflation
         will occur when the symmetry breaking scale associated with the
         defect formation is in the Planck regime
         \be \eta \geq M_{p} \ee

         The criterion for inflation stated above carries over to vortices and
         monopoles as well. Another
         important point is that once started, topological inflation never
         ends. Although the field $\phi$ is driven away from the maximum of
         the potential, the inflating core of the defect, from topology, cannot
         disappear, unless the field unwinds. It has been shown that the core
         thickness grows exponentially with proper time \cite{ruth}. It is also
         worth
         noting that these cases have been displayed robustely in numerical
         simulations.

         We are now equipped to address the issue of defect driven inflation
         in the context of brane-antibrane annihilation in superstring theory.

         \section{Non-Bps D-brane Inflation Scenario: Set Up}
         In the previous section we provided robust conditions for topological
         defects to inflate provided that the core radius is larger than the
         inverse
         Hubble radius. Generically, field theories are difficult to
         understand in the Planck regime, so topological inflation is difficult
         to realize in this context.

         We first wish to shortly discuss the assumptions we are making. We
         will begin by coupling the world volume action for the unstable brane
         system, including the anomaly cancelling terms in the gravity action. Since we
         will only investigate the evolution of the massless degrees of freedom with
         respect to the 6D Planck scale, we shall use an effective gravity in 5+1
         dimensions. Hence the
         6-d Newton constant is
         \be G^{6}=\frac{(\alpha')^{4}g_{s}^{2}}{V(T^{4})} \ee
         where $V(T^{4})$ is the volume of the compact four torus.

         We are now in a position to make a simple consistency check by
         solving for the size of the compactified dimensions in terms of the
         string length scale and coupling constant. As stated in the previous
         section the universal condition to obtain topological inflation is
         that the symmetry breaking scale on the order of the Planck mass.
         \be \eta \sim M_{p} \ee
         From the tachyon potential, the symmetry breaking scale is the string
         length \cite{moore},
         \be \eta = l_{s}^{-1} \label{pl}\ee
         while the six dimensional Planck mass is
         \be M_{pl}=G_{6}^{-1/4} \label {ma}\ee
         Equating \ref{pl} with \ref{ma} we obtain a bound for the size of the
         compactified volume
         \be V(T^{4}) \geq l_{s}^{4}g^{2} \ee
         Hence, for weak string coupling our effective gravity description is
         consistent with our compactification. In other words, the vortex has
         the sufficient thickness in order to undergo inflation.

         How does the tachyon couple to our gravitational 
         action?  While this issue is still under investigation, we 
         provide the following argument\cite{devecchia,oz}.  The tachyon is naturally incorporated into
         boundary string field theory when one rewrites the $U(1)$ field
         strength as a supercurvature \cite{kraus}.
         \be \cal{F}\rm= d\cal{A}\rm \ee
         where
         \be i\cal{A}\rm =
         \left (\begin{array}{cc}
         iA^{+} & \bar{T} \\
         T & iA^{-}

         \end{array} \right)
         \label{SCUR}\ee
         In the Wess-Zumino term the supercurvature has
         $\alpha'$ as a coupling constant.
         \be \int_{M} C\wedge Str e^{2\pi i\alpha'\cal{F}\rm} \ee
         From eq \ref{SCUR}, the tachyon also has $\alpha'$
         coupling and will couple to gravity via. the energy momentum of the $U(1)$
         gauge
         fields in the DBI action . Implicit in this assumption is the
         observation that the time scale for the vortex configuration to
         form is much smaller that the 6D Planck time scale,
         $t_{vortex}<<t^{pl}_{6D}$ This physically means that the tachyon, in
         forming a stable vortex configuration,
         is able to wind around the vacuum manifold to acquire the vortex
         charge faster than the
         gravitational field can backreact to anisotropies of the tachyon field
         dynamics. Hence we can use Birkhoffs theorem to construct a general
         metric solution. For generality, though, we will the tachyon field
         will be time dependent,
	 \be T=T(t) ,\ee
	 even after a stable vortex forms.
         This will be important for the issue of ending inflation.

         Our system will be investigated with the following action
         \be S_{Tot} = S_{Grav} + S^{D-\bar{D}}_{DBI} + S_{WZ}^{D-\bar{D}} \ee
         
         Before proceeding with the explicit calculation it is worth presenting as
         clearly as possible a physical picture in analogy with potential driven
         inflationary
         scenarios. First, we place the tachyon on the same footing as an
         inflaton since it is a scalar field which rolls down a potential. 
	 The second crucial assumption is that the potential of
         the tachyon couples minimally to gravity. Nonetheless, the second
         assumption can be evaded to include non minimal coupling, which has
         also demonstrated topological inflation, but this issue shall
         not be covered in this paper.

         Consider now a $5-\overline{5}$ annihilation. As the five
         branes annihilate, a non-bps 3-brane is created. The creation of this
         3-brane is important as it will act as the spacetime that will
         inflate. The crucial point here is that the core of the vortex configuration
         is
         localized on the whole 3-brane world volume. At the center of the core (in
         this case the
         3-brane) the symmetry is
         restored and the tachyon field vanishes.
         As a result the vortex always remains at the top of the tachyon effective
         potential
         at $T=0$. The false vacuum energy $V(T=0)$ yields a negative pressure
         equation of state for the tachyon field and will drive an inflationary
         epoch of the 3-brane world volume. The crucial point which differs
         from ordinary inflation is that this mechanism dynamically tunes the
         tachyon potential to the optimal value for inflation on the brane by
         localizing all of the vacuum energy on a (3+1)D hypersurface.

         \section{Vortex Inflationary Solution}

         Tachyonic condensation in the $D-\bar{D}$ system
         flows from the false open string vacua to the closed string vacuum.
         In our case, there is a remnant $D(p-2)$-brane formed after the tachyonic
         field winds nontrivially around the vacuum
         manifold.   In the $10 D$ target space the core of this defect is indeed the

         world-volume of the $Dp-2$ brane and hence has trapped false vacuum
         energy from the non-trivial winding of the tachyon. Since the system is in
         the closed string vacuum, where gravitational
         interactions are turned on, this vortex carries energy-momentum.
         Although it is not well understood how gravity is
         incorporated into the tachyon condensation proccess, we shall 
         argue by
         the consistency of the six-dimensional coupling constant in the
         gravitational sector and the perturbative effective field theory
         describing the energy-momentum of the vortex brane.

         We are assuming that the
         tachyonic field has formed the vortex configuration before the
         gravitational interactions are turned on. This is consistent with the
         annihilation proccess since the system first begins in an open string
         false vacuum state and evolves to a closed vacuum. Hence, the defect
         configuration will impose a most general
         form
         of the time dependent metric.

         The effective action we will study is

         \be S= S_{gravity} + S_{vortex} + S_{WZ}\ee

         \be S= -\frac{2}{G_{6}}\int d^{6}x \sqrt{-g}[(R-2\Lambda)
         -2G_{6} \cal{L}\rm_{DBI} + \cal{L}\rm_{WZ}], \ee
         where R is the 6-dimensional scalar curvature,$\cal{L}\rm_{DBI}$ is the
         complete $D-\bar{D}$ Lagrangian and $G_{6}$ is the six dimensional Newton
         constant. In particular, the tree level effective
         Lagrangian for the tachyonic field is
         \be \cal{L}\rm_{DBI} =
         \frac{1}{g^{2}_{YM}} \int dx^{6}\sqrt{-g}
         \left[F_{\mu\nu}^{\pm}F^{\pm \mu\nu} -
         (D_{\mu}T)^{2} - \frac{1} {2}\rm (|T|^{2} -\Psi^{2})^{2}\right] \ee
         and the Wess-Zumino term
         \be \cal{L}\rm_{WZ} =T_{D5}\int_{M_{6}} C\wedge Str e^{2\pi
         i\alpha'\cal{F}\rm}, \ee
         where the supertrace 
	 \be Str M = Tr(-)^{F} M= Tr\left(
         \begin{array}{cc}
         1 & 0 \\
         0 & -1
         \end{array} \right)M \ee

         For the $D5-\bar{D}5$ problem the Wess-Zumino term becomes
         \be
         T_{D5}\int_{M_{6}} C_{4}\wedge (2\pi\alpha')dT\wedge d\bar{T} \ee

         Upon varying the above action we obtain the 6D Einstein Equations
         \be G^{6}_{\mu\nu}= G_{6} T_{\mu\nu}, \ee
         where the energy-momentum tensor of the $D5-\bar{D}5$ system is
         \be T_{\mu\nu}=\frac{\delta{S}}{\delta g^{\mu\nu}} =
         D_{\mu}T\bar{D}_{\nu}\bar{T} + D_{\nu}T\bar{D}_{\mu}\bar{T} -
         g^{\alpha\beta}F^{-}_{\mu\alpha}F^{-}_{\nu\beta} + g_{\mu\nu}\cal{L}\rm
         \ee
         The tachyon E.O.M is
         \be D_{\mu}D^{\nu}T = \frac{\partial V(T\bar{T})} {\partial\bar{T}} \ee
         where $V(T\bar{T})$ is the tachyon potential in eq 
         (\ref{tpot}),  while the E.O.M for the gauge field is
         \be \nabla^{\nu}F^{-}_{\mu\nu}=ie(\bar{T}\nabla_{\mu}T -
         T\nabla_{\mu}\bar{T}) - 2e^{2}A^{-}_{\mu}T\bar{T} . \ee

         The general time dependent tachyon vortex solution is
         \be T(t,r)= \phi(t,r)e^{in\theta} \ee
         \be A^{-}_{\mu} = \frac{n}{e}\beta(t,r)\nabla_{\mu}\theta \ee
         subject to the energy conserving boundary conditions
         \be \begin{array}{cc}
         \phi(t,0)=0  & \phi(t,\infty)=\Psi, \\
         \beta(t,0)=0  &  \beta(t,\infty)=1.
         \end{array} \ee

         The vortex configuration is localized on the
         co-dimension 2 hypersurface, identified as a D-3 brane in our case.
         The most general solution is:

         \be ds_{5+1}^{2}= g_{\mu\nu}dx^{\mu}dx^{\nu} + g_{ij}dx^{i}dx^{j} \ee
         where $g_{\mu\nu}$ and $g_{ij}$ are the brane and transverse metrics
         respectively. The general time dependent solution which satisfies the
         Einstein field equations with planar symmetry in five space-time direc		tions is:
         \begin{eqnarray} &ds_{5+1}^{2} &=  -dt^{2} + B(t,r)^{2} dr^{2}
         + H(t,r)^{2}(dx_{1}^{2} + dx_{2}^{2} + dx_{3}^{2}) + \nonumber \\
	 & & C^{2}(r,t)r^{2}{d\theta}^{2} \end{eqnarray}
	 Then the tachyon equation of motion becomes
	  \be \begin{array}{l} \ddot{T} + (\frac{\dot{B}}{B} - \frac{\dot{C}}{C}
         -\frac{\dot{H}}{H})\dot{T} \\
         + \frac{T''}{B^{2}} + \frac{1}{C^{2}r^{2}}T(1-\alpha)^{2}  + T(T^{2}-
         \psi^{2})=0 \end{array} \ee
         where $'$ denotes $\partial_{r}$.
        
	 We shall now proceed to solve for the metric coefficients and look
         for inflating solutions specifically of the co-dimension 2
         hypersurface; the 3-brane world volume. This solution describes a
         localized 3-brane sourced by the false vacuum energy of the tachyonic
         vortex , whose core lives on the
         3-brane worldvolume.  The tachyon vanishes at the core, thus satisfying the condition
         for defect driven inflation.  It has been demonstrated that, in 
         this case, both
         the 3-brane
         and the transverse coordinates will undergo exponential inflation
	 \cite{ruth,vilen3,cho}.

         We will now discuss two separate cases of the space-time solution:

         (1) The gauge field $A^{-} =0$

         (2) The gauge field $A^{-} \neq 0$

         In the latter case the field equations are difficult to solve
         analytically for all times,$t$, since the gauge field is also time
         dependent. However,
         around the center of the vortex
         \be \frac{\dot{B}}{B} = \frac{\dot{C}}{C} = \frac{\dot{H}}{H} =
         \sqrt{\frac{8\pi G}{3}V(T=0)} . \ee

         We immediately see that inflation occurs along the 3-Brane world
         volume as well as the transverse directions.

         When the gauge field is set to zero the solutions correspond to a
         global vortex which has been shown by other authors to exhibit a
         warped geometry with de-Sitter expansion of the 3-brane world volume
         directions\cite{vilen3}.
         It was shown that the general solution interpolates between a $dS_6$ and
         a $dS_{4}\times R_{2}$.

         \section{Curved Vortex: How Inflation Ends and Matter Remains}
         In earlier versions of topological inflationary scenarios, the
         inflaton remains at the maximum of the false vacuum yielding eternal
         inflation. Therefore, there is no end to inflation once it sets in.
         This phenomenon occurs by virtue of the no hair theorem for de-Sitter
         space: In the core of the defect ($T=0$) the space-time evolves by
         its own laws and continues to expand exponentially at all times.

         Eternal inflation will occur in our case as well. However, there is a
         possible way to end inflation quite naturally in the brane case.
         The vortex lives in a transverse two dimensional hypersurface and
         has the associated translational invariance. It was shown that the
         fluctuations of the tachyon field on the vortex are the collective
         coordinates of the vortex\cite{arkady2}. If the vortex moves in the
         bulk transverse direction, the tachyon field can unwind. 

         Once the five branes annihilate, the vortex forms about the three
         dimensional hypersurface in the bulk. However, this vortex will in general have an extrinsic curvature due to its
         embedding in
         the curved bulk. We wish to make an analogy at this stage with the physics

         of cosmic strings defined by the Nambu Goto action. The extrinsic
         curvature of cosmic strings gives rise to its relativistic
         velocity along the direction normal to the length of string \footnote{I
         thank Robert Brandenberger for making this important 
         connection.}.  Similarly, the nucleated vortex in our case will also be curved and
         possess
         a non-zero velocity away from the initial position where it was nucleated.
         Hence, once
         inflation sets in on a given space-time 
         hypersurface,$\cal{M}\rm_{infl}$,
         it will eventually end
         as the vortex moves away from $\cal{M}\rm_{infl}$ where inflation
         was initiated [see fig2]. This occurs because
         the time dependent tachyon field leaves the top of the false vacuum.
         The vacuum energy decreases in our space-time and the expansion rate
         will decrease the further away the center of the vortex is from our
         initial space-time hypersurface.

         One might be worried that the confined U(1) gauge field localized on the
         vortex
         will leave the hyperfurface that is identified as our space-time once
         the vortex moves away; while we
         want matter to remain on our space-time. But since our vortex is
         relativistic there can be relativistic self intersection (cusps) along the

         vortex \footnote{I am indebted to Joao Magueijo for pointing this out to
         me.}. It has been demonstrated \cite{blanco} that self intersecting
         chiral vortices shall emit loops of superconducting cosmic
         strings\cite{witten},
         namely vortons. This could be a viable mechanism for generating matter
         and density perturbations in the early universe and is worth further
         investigations. [For a review see \cite{vilenkin,hindmarsh}
         \begin{figure}
         \centerline{\epsfxsize=3in\epsffile{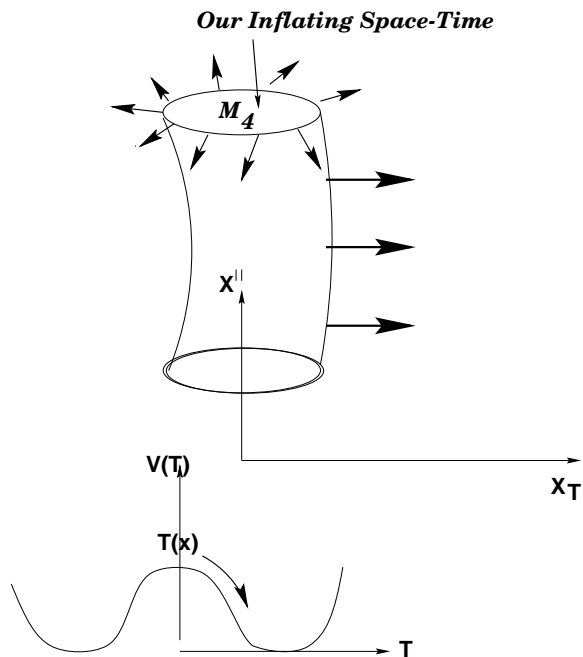}}
         \caption{The vortex has extrinsic curvature and will move away from the
         initial inflating hypersurface it was created on. This causes the
         tachyon field to roll down the false vacuum potential eventually ending
         inflation on the hypersurface $\cal{M}\rm_{4}$}
         \label{w}

         \end{figure}

         \section{Conclusions and Discussion}

         When formulated in
         conventional quantum field theory coupled to gravity, inflation exhibits
         initial condition fine tuning, singularity and trans-Planckian problems.
         We have suggested
         a dynamical inflationary mechanism resulting from 
         $D-\bar{D}$ brane annihilation to address these problems.  These branes are in
         the non-BPS sector of superstring theory.  This
         mechanism is analogous
         to a ``Big-Bang'' mechanism, in that the branes hit each other, annhilate
         and a lower dimensional inflating brane emerges as result of the
         annnihilation proccess.
         Moreover, we have made a concrete connection with Vilenkin's and
         Linde's realization of topological inflation.  In both 
         models, there is little need of fine tuning of potentials.
         In our model, the tachyon condensate forms a vortex
         whose core is localized on the $D3$-brane world volume which sets the
         initial condition neccessary for inflation.

         Since our scenario resembles topological inflation the question of
         how inflation will end becomes relevant. We argue that inflation will end
         naturally due to the motion of vortex away from the space time
         hypersurface within which inflation was initiated. This occurs 
         because the tachyon field defined at the initial hypersurface is a collective
         coordinate of the vortex and will minimize the vacuum energy as it
         rolls down the potential which is associated with the motion of the
         vortex in the bulk away from our initial inflating hypersurface.

         In light of the stringy inflationary mechanism presented in this
         paper, one is led to a few outstanding puzzles. First, string
         theory possesses a myriad of D-brane species and this mechanism could
         , in principle, apply to other inflating hypersurfaces of differing
         dimensionalities. Is there something unique about an inflating 3+1
         D hypersurface, resulting from annihilating Dp-branes?
         Non-perturbative data from string theory should shed new light on
         this question and we leave this issue for future investigations.
         Interestingly, a similar 'big-bang' scenario avoiding an inflationary
         epoch has been
         suggested by colliding branes in the context of Horava-Witten
         compactification \cite{turok}.

         There is a stringy mechanism which selects a large 3+1D 
         space-time from annihilating D1-branes, namely the Brandenberger-Vafa scenario
         (BV)\cite{bv,abe}. This situation takes the annhiliation of a D(p-2) brane
         into a large p spatial dimenison for p=3. If in the BV mechanism we
         associate this large dimension as a ``brane world'' then there is a
         similarity in that our mechanism takes the annihilation of a
         Dp-brane into an inflating D(p-2) space for p=5. These two pictures
         are intriguingly related to each other via. Myer's dielectric
         effect\cite{rob} which has been employed to resolve 
         gravitational naked singularities \cite{cliff}. We believe
         that further investigation of this issue will be illuminating.

         \acknowledgements
         I wish to give special thanks to Emil Akhmedov, Robert Brandenberger, Jussi Kalkkinen
         Sanjaye Ramgoolam and Arkady Tseytlin for illuminating discussions. I am also thankful to Clifford Johnson, Joao
         Magueijo, Matty Parry, Ashoke Sen, Kelly Stelle and Radu Tatar for
         discussions and inspiration.

         \end{document}